\documentclass[12pt]{article}
\newcommand{\xxx}[1]{ [#1]}
\newcommand{\mysection}[1]{\section{#1}
   \hspace{0.8cm}\setcounter{equation}{0}}
\renewcommand{\theequation}{\arabic{section}.\arabic{equation}}
\newcommand{\myappendix}{\appendix
   \renewcommand{\theequation}{\Alph{section}.\arabic{equation}}
   \vspace{30pt} \noindent {\Large \bf Appendices}}
\newlength{\dummysp}
\newlength{\bdummysp}
\newcommand{\la}{\lambda}
\def\|{|_{(0,0)} \, }
\newcommand{\beq}{\begin{equation}}
\newcommand{\eeq}{\end{equation}}
\newcommand{\ben}{\begin{enumerate}}
\newcommand{\een}{\end{enumerate}}

\newcommand{\anneq}[1]{\mathop{=}\limits_{#1}}

\newcommand{\tr}{\mathop{{\hbox{tr} \, }}\nolimits}

\newcommand{\stxt}[1]{\mathop{\hbox{{\scriptsize #1}}}\nolimits}

\newcommand{\bbar}[1]{{\overline{#1}}}
\newcommand{\spc}{\hbox{\hspace{\dummysp}}}
\newcommand{\bspc}{\hbox{\hspace{\bdummysp}}}
\newcommand{\half}{{1 \over 2}}

\newcommand{\third}{{1 \over 3}}
\newcommand{\tthird}{{2 \over 3}}

\newcommand{\beqa}{\begin{eqnarray}}
\newcommand{\eeqa}{\end{eqnarray}}
\newcommand{\nnn}{ \nonumber \\ }

\newcommand{\p}{{\partial}}
\newcommand{\W}{{\cal W}}
\newcommand{\Wb}{{\bbar{\cal W}}}
\newcommand{\D}{{\cal D}}
\newcommand{\Db}{{\bbar{\cal D}}}

\newcommand{\e}{{\epsilon}}
\newcommand{\s}{{\sigma}}
\newcommand{\sbar}{{\bar \sigma}}
\newcommand{\adot}{{\dot \alpha}}
\newcommand{\bdot}{{\dot \beta}}

\newcommand{\vev}[1]{{\langle #1 \rangle}}

\newcommand{\ord}[1]{{{\cal O}(#1)}}

\newcommand{\gappeq}{\mathrel{\rlap {\raise.5ex\hbox{$>$}}
{\lower.5ex\hbox{$\sim$}}}}
\newcommand{\lappeq}{\mathrel{\rlap{\raise.5ex\hbox{$<$}}
{\lower.5ex\hbox{$\sim$}}}}
\newcommand{\myref}[1]{(\ref{#1})}
\newcommand{\dx}{{\delta_X}}

\newcommand{\ove}{{1 \over e}}
\newcommand{\Lcal}{{\cal L}}

\newcommand{\lbar}{{\bar \ell}}
\newcommand{\Fbar}{{\bbar{F}}}
\newcommand{\phibar}{{\bar \phi}}
\newcommand{\phib}{{\bar \phi}}
\newcommand{\Dbf}{{\bf D}}

\newcommand{\fourth}{{1 \over 4}}
\def\[{\left[}
\def\]{\right]}
\def\({\left(}
\def\){\right)}
\newcommand{\sint}{\int d^4 \theta \,}
\newcommand{\hc}{{\rm h.c.}}
\newcommand{\Phibar}{{\bbar{\Phi}}}

\newcommand{\sqtw}{\sqrt{2}}
\newcommand{\psibar}{{\bar \psi}}
\newcommand{\psib}{{\bar \psi}}
\newcommand{\Mbar}{{\bbar{M}}}
\newcommand{\Mb}{\Mbar}
\newcommand{\ubar}{{\bbar{u}}}
\newcommand{\ub}{{\bar u}}

\newcommand{\kbar}{{\bar k}}
\newcommand{\mbar}{{\bar m}}
\newcommand{\Gkkb}{G_{k \kbar}}
\newcommand{\Dau}{\D^\alpha}
\newcommand{\Dad}{\D_\alpha}
\newcommand{\Dbadu}{\Db^\adot}
\newcommand{\Dbadd}{\Db_\adot}

\newcommand{\uone}{$U(1)$}
\newcommand{\ap}{{(a)}}
\newcommand{\WWa}{(\W \W)_\ap}
\newcommand{\lla}{(\la \la)_\ap}
\newcommand{\ha}{h^\ap}
\newcommand{\Ua}{{U_\ap}}
\newcommand{\Uba}{{{\bbar{U}}_\ap}}
\newcommand{\ua}{{u_\ap}}
\newcommand{\hha}{{\hat h}^\ap}
\newcommand{\sumC}{\sum_{\ap \in G_C}}
\newcommand{\sumNC}{\sum_{\ap \not\in G_C}}
\newcommand{\hr}{{\hat r}}
\newcommand{\Pk}{{\Phi^k}}

\newcommand{\pk}{{\phi^k}}
\newcommand{\Pbk}{{\Phibar^\kbar}}
\newcommand{\pbk}{{\phibar^\kbar}}

\newcommand{\Ds}{\D^2}
\newcommand{\Dbs}{\Db^{\hspace{1pt} 2}}
\newcommand{\WZ}{\anneq{\stxt{WZ}}}
\newcommand{\Dbbdd}{\Db_\bdot}
\newcommand{\Dbd}{\D_\beta}
\newcommand{\ttz}{$\theta = \bar \theta = 0$}
\newcommand{\eig}{{1 \over 8}}
\newcommand{\iotw}{{i \over 2}}
\newcommand{\vp}{\varphi}
\newcommand{\vpb}{{\bar \varphi}}
\newcommand{\vpbs}{(\vpb \vpb)}

\newcommand{\vps}{(\vp \vp)}

\newcommand{\La}{\Lambda}
\newcommand{\Lb}{{\bar \Lambda}}
\newcommand{\lb}{{\bar \lambda}}
\newcommand{\tWW}{\tr(\W\W)}
\newcommand{\tWWb}{\tr(\Wb\Wb)}
\newcommand{\tll}{\tr(\la \la)}

\newcommand{\Gaad}{G_{\alpha \adot}}

\newcommand{\Baad}{B_{\alpha \adot}}
\newcommand{\Rb}{{\bbar{R}}}
\newcommand{\Fk}{F^k}
\newcommand{\Fbk}{\Fbar^\kbar}

\newcommand{\chiproj}{(\Dbs-8R)}
\newcommand{\chiprojb}{(\Ds-8\Rb)}
\newcommand{\tfchiproj}{(\Dbs-24R)}
\newcommand{\tfchiprojb}{(\Ds-24\Rb)}
\newcommand{\chib}{{\bar \chi}}
\newcommand{\vpa}{\vp_\alpha}
\newcommand{\vpau}{\vp^\alpha}

\newcommand{\ckl}{(\chi^k \chi^\ell)}
\newcommand{\cbkl}{(\chib^\kbar \chib^\lbar)}
\newcommand{\chiak}{\chi_\alpha^k}

\newcommand{\sixth}{{1 \over 6}}

\newcommand{\ddd}{\nnn &&}
\newcommand{\Ub}{{\bbar{U}}}
\newcommand{\chibk}{\chib^\kbar}
\newcommand{\chik}{\chi^k}
\newcommand{\phibk}{\phib^\kbar}
\newcommand{\phik}{\phi^k}
\newcommand{\Rbar}{{\bbar{R}}}

\newcommand{\nUcb}{{\( \ub - (\lb^\ap \lb_\ap) + {4 \over 3} \Mb \ell \)}}
\newcommand{\nUc}{{\(  u - (\la^\ap \la_\ap) + {4 \over 3} M \ell \)}}
\newcommand{\hf}{{\hat f}}
\newcommand{\ux}{$U(1)_X$}
\begin{document}

\begin{titlepage}

\baselineskip=18pt

\renewcommand{\thefootnote}{\fnsymbol{footnote}}

\hfill    LBNL-50327

\hfill    UCB-PTH-02/22

\hfill    hep-th/0205206

\hfill    Mar.~27, 2003

\vspace{20pt}

\begin{center}
{ \bf \Large Full Component Lagrangian \\
in the Linear Multiplet Formulation \\

\vskip 5pt

of String-inspired Effective Supergravity}
\end{center}

\vspace{20pt}

\begin{center}
{\sl Joel Giedt}\footnote{E-Mail: {\tt giedt@physics.utoronto.ca}}
\end{center}

\vspace{20pt}

\begin{center}
{\it Department of Physics, University of California, \\
and Theoretical Physics Group, \\
Lawrence Berkeley National Laboratory, Berkeley, \\
California, U.S.A.}\footnote{This work was supported in part by the
Director, Office of Science, Office of High Energy and Nuclear
Physics, Division of High Energy Physics of the U.S. Department of
Energy under Contract DE-AC03-76SF00098 and in part by the National
Science Foundation under grant PHY-00-98840.}

{\it and}

{\it Department of Physics, University of Toronto, \\
60 Saint George Street, Toronto, \\
Ontario\hspace{2pt} M5S 1A7 \hspace{2pt} Canada}\footnote{Support
was also provided by NSERC.}
\end{center}

\vfill

\end{titlepage}

\renewcommand{\thepage}{\roman{page}}
\renewcommand{\thefootnote}{\arabic{footnote}}
\setcounter{page}{2}
\mbox{ }

\baselineskip=18pt

\vskip 1in

\begin{center}
{\bf Abstract}
\end{center}

\vspace{30pt}

\noindent
We compute the component field
4-dimensional $N=1$ supergravity Lagrangian
that is obtained from a superfield Lagrangian
in the $U(1)_K$ formalism with a linear
dilaton multiplet. All fermionic terms
are presented.  In a variety of important
ways, our results generalize
those that have been reported previously,
and are flexible enough to
accomodate many situations of phenomenological
interest in string-inspired effective supergravity,
especially models based on orbifold
compactifications of the weakly-coupled heterotic string.
We provide for an effective theory of
hidden gaugino and matter condensation.
We include supersymmetric Green-Schwarz counterterms associated
with the cancellation of \uone\ and modular
duality anomalies; the modular duality counterterm
is of a rather general form.
Our assumed form for the dilaton K\"ahler potential is
quite general and can accomodate K\"ahler stabilization
methods.  We note possible applications of our results.
We also discuss the usefulness of the linear dilaton
formulation as a complement to the chiral dilaton approach.

\newpage
\renewcommand{\thepage}{\arabic{page}}
\setcounter{page}{1}
\setcounter{footnote}{0}

\mysection{Introduction}
The topic of {\it 4-dimensional $N=1$ supergravity}
coupled to supersymmetric matter and super-Yang-Mills fields is an
old and well-understood part of supersymmetric
field theory.  A variety of superspace methods
were developed many years ago, all of them designed
to write locally supersymmetric Lagrangians
in a compact form while at the same time leading
most easily to component field expressions.
Nevertheless, this {\it component expansion}
can be tedious and it proves useful
to have the results tabulated; e.g., so that model-building
can proceed, with a minimum of effort,
from superfield assumptions
to component field consequences.

For the case of chiral multiplets coupled
to supergravity and the vector supermultiplets
of a super-Yang-Mills theory,
component expansions have been
tabulated under a very broad set assumptions
(originally by Cremmer et al.~\cite{CFGVP}; nicely reviewed
by Wess and Bagger~\cite{WB92}).
For less conventional arrangements and assortments
of $N=1$ multiplets, however,
the coverage is a bit patchy.
The example with which we are interested involves
a {\it linear multiplet}~\cite{FZW74}.
Although supergravity coupled to a linear
multiplet~\cite{Sie79} has been studied in a large number
of works (enumerated and discussed below), some
details remain to be given.  This is particularly true
when very general assumptions for the form
of the superfield Lagrangian are envisioned.

In the present work we consider a class of
supergravity theories---containing a linear
multiplet---for which only specialized
or somewhat incomplete results are available;
it is our intention to generalize previous work
and to fill in some details that are missing
in the literature.

While completeness is a reasonable motivation for the
determination of the full component Lagrangian, Planck
mass suppressed fermion interactions may have phenomenological
applications.  For example, processes forbidden in
the Standard Model and globally supersymmetric extensions
might be mediated by gravitationally suppressed interactions.
This is particularly true in the case where large
vacuum expectation values ({\it vev}s)
occur due to the presence of an anomalous \uone.

Below, we will argue that a theory of supergravity that
contains a linear multiplet is well-motivated from the
perspective of {\it string-inspired effective supergravity.}
Our discussion is a synopsis of opinions offered previously
by other authors.  Whereas the {\it chiral dilaton} formulation
is more common, on the basis of the points raised
in our discussion it is our opinion that the {\it linear dilaton}
framework should be regarded as a useful complement.

We are interested in generalizations of the effective
theory of Bin\'etruy, Gaillard and Wu
(BGW)~\cite{BGW96,BGW97a,BGW97b}, as well
as in the computation of those fermionic terms in the Lagrangian
that were neglected by these authors (justifiably,
as the issues in which they were interested
did not require knowledge of these terms).  The {\it BGW
effective theory} is inspired by
orbifold compactifications of the weakly-coupled
heterotic string.\footnote{More specifically,
BGW were concerned with the $E_8 \times E_8$
heterotic string.  However their effective
supergravity description would work just as well for
$spin(32)/Z_2$ constructions.  The only issue
there is to {\it hide} the {\it hidden sector.}
Of course this is already a problem in the
$E_8 \times E_8$ case due to the presence of {\it twisted
sectors,} which couple subgroups of the two $E_8$'s;
for example, see the discussion in \cite{CKM89,Gie02a}.}
The low-energy limit of
the heterotic string compactified on an orbifold
is an effective supergravity theory.
The BGW effective theory is designed to implement
dynamical supersymmetry breaking through the
strong dynamics of a super-Yang-Mills theory
in a {\it hidden sector.}  Whereas the perturbative
scalar potential of the effective supergravity
has flat directions---corresponding to an
infinite vacuum degeneracy parameterized by
massless scalars ({\it moduli})---the
effective theory of dynamical supersymmetry breaking
lifts these flat directions and hence gives
rise to {\it moduli stabilization.}

In Section \ref{eft} we outline the field content
that is present in the theories studied here.
We describe the interpretations that are to be given
to these fields in the context of string-inspired
effective supergravity.  We briefly review the
{\it duality} that relates the
chiral dilaton and linear dilaton formulations.
We discuss the reasons why one might choose to
work in the latter formulation in addition to the former.  We
comment on some instances in
which the two formulations have confronted
each other and seem to be at odds.  A brief summary of previous
works on the subject of supergravity with
a linear dilaton is given.  We relate these to our work
and describe how our results supplement
those that already exist.  Finally,
we define the superfield Lagrangian for
which we have computed the component expansion.

Section \ref{resu} discusses some aspects of our results.
We emphasize features of the component
Lagrangian that we find interesting.
We guide the reader
to our main results---{\it lengthy formulae}---which
are contained in appendices.
Our conclusions are stated in Section~\ref{conc}.

In Appendix \ref{cna} we summarize our notations,
conventions and abbreviations.
In Appendix \ref{proj} we outline the method of
{\it projection} to component fields that is used
in the $U(1)_K$ superspace formalism (see below).
In Appendix \ref{geid} we discuss the geometric
identities of $U(1)_K$ superspace that are particularly
useful in the computation of the component Lagrangian.
In Appendices \ref{ezpt}-\ref{mgse} we present the
lengthy formulae that comprise our principle results.

\mysection{The effective theory}
\label{eft}
In this section we introduce the reader to the class
of theories that we intend to study.
We relate our approach
to the more familiar formulation with a chiral dilaton.
We also discuss the motivations for working with
a linear dilaton.  Indeed, we believe
that there exist instances where one might
prefer to use a linear dilaton; these
reasons are of a purely {\it practical}
nature, as the two formalisms are equivalent
provided they are {\it properly} related.
(Specifically, the equations of motions should
form equivalent systems and
constraint equations in one formalism must
find an equivalent expression in the other.)

\subsection{Content and framework}
Our intent in this subsection is only to present
enough of a summary that the reader not familiar
with these topics can understand the motivations
and key concepts involved in the present work.
For further details the reader is in invited to
refer to the articles that we cite below;
in particular the reviews \cite{GG99,BGG01} are valuable
for further study of formal issues of
linear multiplets in the $U(1)_K$ formalism
and Refs.~\cite{Wu97,Gai98}
provide nice reviews of the related superstring
phenomenology.

The theory we consider is a variety of four-dimensional
$N=1$ supergravity.  It contains the graviton, Yang-Mills gauge fields,
matter fields and moduli fields, including a dilaton.
Each of the matter and moduli fields (excepting
the dilaton) may or may not
be charged with respect to the Yang-Mills
gauge group.  All of the fields in the theory are introduced
along with superpartners through $N=1$ supermultiplets,
by starting with the {\it K\"ahler $U(1)$ superspace}
formalism (denoted $U(1)_K$) \cite{BGGMa,BGGMb,ABGG93}.
The $U(1)_K$ approach has been reviewed in \cite{BGG01},
{\it hereafter referred to as BGG.}
The minimal supergravity multiplet
$(e_m^{\spc a}, \psi_m^{\spc \alpha}, \psib_{m \adot}, M, \Mb, b_a)$
is introduced through the superdeterminant $E$ of the supervielbein
$E_M^{\bspc A}$ and geometric relations in $U(1)_K$ superspace.
Gauge multiplets
$(a_{\ap m}, \la_{\ap \alpha}, \lb_\ap^\adot, \Dbf_\ap)$
are introduced through vector superfields fixed to
Wess-Zumino gauge,
where $\ap$ labels a basis of orthogonal generators
of the gauge group. Except in a counterterm
associated with an anomalous \uone\ factor,
only corresponding chiral
field strengths $\W^\alpha_\ap$ will appear explicitly
in the superfield Lagrangian, due to the appearance
of the Yang-Mills connection in the covariant
derivatives of $U(1)_K$ superspace.
Matter multiplets and all of
the moduli multiplets except the dilaton are introduced through
chiral superfields $\Phi^k$, which have field content
$(\phi^k, \chi^k_\alpha, F^k)$.  The dilaton
is introduced through a (modified) linear superfield $L$;
its field content will be discussed below.
We account for the leading effects of a strongly coupled
{\it hidden sector} with condensing gauge group $G_C$
through static (auxiliary) chiral superfields
$\Ua$ and $\Pi^\alpha$.  Here, $\Ua$ is in correspondence
with the operator $\WWa$ in the unconfined theory; i.e.,
its lowest component (the \ttz\ part, denoted by
$\|$) $u_\ap = U_\ap \|$
corresponds to the gaugino bilinear operator $\lla$ which
acquires a nonvanishing vev ({\it vev}), triggering
{\it gaugino condensation} \cite{gcd}.  The lowest component of $\Pi^\alpha$
corresponds to a scalar operator of
hidden sector matter fields which may also take a nonvanishing vev
and play a role in the effective theory
of supersymmetry breaking through the dynamics of the
hidden sector.

\subsection{Linear versus chiral dilaton}
In the context of string-inspired supergravity,
the 4-dimensional dilaton is a composite of
10-dimensional fields dimensionally reduced to an effective
four-dimensional theory.  In Witten's classic
reduction \cite{Wit85}, one has
a real scalar $\s$, which arises from the 10-dimensional
graviton, the 10-dimensional dilaton $\phi$, and a
2-form field strength $\hat f_{mnp}$.
The 4-dimensional dilaton is given by
\beq
{1 \over g_s^2(\s,\phi) } = e^{3\s} \phi^{-3/4}.
\eeq
It is the vev
of this quantity that determines
the strength of gauge couplings.  The effective
supergravity Lagrangian naturally pairs this
4-dimensional dilaton with a pseudoscalar $D$ that
is the universal axion; however, a duality transformation must
be made to trade $\hat f_{mnp}$
for $D$:
\beq
\phi^{-3/2} e^{6\s} \hat f_{mnp} \equiv \e_{mnpq} \p^q D.
\label{cfdt}
\eeq
Then the natural pairing is
\beq
s = e^{3\s} \phi^{-3/4} + 3i \sqtw D,
\eeq
since for a K\"ahler potential with leading order $s$ dependence
$K \ni - \ln(s + \bar s)$ the standard chiral supergravity
formulation yields the correct terms in the effective
Lagrangian.  When we promote $s \to S$, a chiral
superfield, we have the chiral dilaton formulation.
Due to the $N=1$
supersymmety, the complex field $s$ has a superpartner
which is the {\it dilatino.}  The chiral multiplet formulation
is used for reasons of familiarity and simplicity.

Instead of making the duality transformation \myref{cfdt},
we can work with an $N=1$ multiplet that already
contains a 2-form field strength---the
linear multiplet \cite{FZW74,Sie79,sjg,BGGa}.  Whereas in the
leading-order effective Lagrangian it
is straightforward to impose \myref{cfdt}
and replace $\hat f_{mnp}$ by $D$,
in a more general setting one finds that the
corresponding duality transformations are difficult to
perform explicitly \cite{DFKZ92,DQQ94,NelPC}.
If the beyond-leading-order Lagrangian is obtained from
string theory, so that it already contains the
2-form field strength $\hat f_{mnp}$, it may be more
practical to work with the linear multiplet and thus
avoid intricacies that may be
associated with the duality transformation.

A second issue arises when an effective theory of gaugino condensation
is included as a mechanism for dynamical supersymmetry
breaking.  The chiral field strength superfields
of the Yang-Mills group satisfy
\beq
\tfchiprojb \tWW - \tfchiproj \tWWb = {\rm t.d.},
\label{tfbi}
\eeq
where ``t.d.'' stands for a total derivative.
To treat the condensate superfield $U \sim \tWW$
as an ordinary chiral superfield (of $U(1)_K$ weight 2)
fails to implement this constraint \cite{BDQQ95,BGT95,BG96}.
In the linear dilaton formulation
the chiral field strength emerges from
the {\it modified linearity conditions:}
\beq
\chiproj L = - \tWW, \qquad \chiprojb L = - \tWWb .
\label{plpj}
\eeq
When $U = \sumC \Ua$ (where $G_C$ is the condensing
part of the gauge group) is introduced in \myref{plpj} through
\beq
\tWW \to \tWW + U,
\eeq
then \myref{tfbi} is {\it automatically satisfied}
for the $\Ua$'s \cite{BGW97a}.
Although this constraint {\it can} be imposed
in the chiral dilaton multiplet formulation,
it is more difficult.  In this regard the linear multiplet
has a practical advantage.

Of course one may ask: (i) what evidence exists
that would suggest \myref{tfbi} {\it should} be satisfied
when $\tWW$ is replaced by the interpolating field $U$;
(ii) whether imposing this constraint has any
important effects on the effective theory of
dynamical supersymmetry breaking.  These
are certainly fair questions and we
know of no clear answer to (i), except that
it seems like the most reasonable assumption.  We
do have something to say about (ii).

In the present formalism the condensate superfields $\Ua$
are introduced as static chiral superfields.  Their
highest components $F_{\Ua}$ (defined in \myref{krrt}) thus
appear only linearly in the component Lagrangian.
However a subtlety arises in deriving the corresponding
equations of motion:
an important constraint exists on the $F_\Ua$
if we extend \myref{tfbi} to the condensates.
That is, suppose we impose
\beq
\tfchiprojb (\tWW + U) - \tfchiproj (\tWWb + \Ub)  = {\rm t.d.}
\label{jllo}
\eeq
Then it was noted in \cite{BGW96} that we have
for the highest components $F_{\Ua}$ the constraints:
\beq
-\fourth \( \Ds \Ua - \Dbs \Uba \)
= F_\Ua - \Fbar_\Uba = 4i \nabla_m B_\ap^m
+ u_\ap \Mbar - \ubar_\ap M .
\label{fdfc}
\eeq
Here $\nabla_m B_\ap^m$ is identified with the
total derivative term in \myref{jllo}, while
$u_\ap \Mbar - \ubar_\ap M$ arises from the
$24 \Rbar U - 24 R \Ub$ part.
When one varies the action with respect to the
auxiliary fields $F_\Ua$,
it is crucial to respect \myref{fdfc} by first
rewriting $F_\Ua$ as
\beq
F_\Ua = \half \(F_\Ua + \Fbar_\Uba\) + 2i \nabla_m B_\ap^m
+ \half \(u_\ap \Mbar - \ubar_\ap M\)
\label{jjhe}
\eeq
and the conjugate of this for $\Fbar_\Uba$.
(For example, this has
been done in Eq.~(2.21) of \cite{BGW97a}.)
One then varies with respect to the
{\it unconstrained} combination
$F_\Ua + \Fbar_\Uba$.

The crucial thing to notice is the last term on the
right-hand side of \myref{jjhe}.  Generically it
has a nonvanishing vev when the scalar potential
is minimized.\footnote{Clearly the phases of
the condensates $u_\ap$ and the auxiliary
scalar $M$ are intimately involved in
whether or not this term vanishes in the
vacuum.  Thus a more detailed study of the axionic
background is necessary to understand its
relevance.  For example, see \cite{BGW97a}.}
It is a supersymmetry breaking vev
which couples to operators that appear with a coefficient
$F_\Ua$ in the Lagrangian.  In particular it
can contribute to soft terms in the low energy
effective Lagrangian.  Note also that $u_\ap \Mbar - \ubar_\ap M$
is anti-Hermitian whereas $F_\Ua + \Fbar_\Uba$ is Hermitian.
Thus these operators couple to different parts of
the operators that are coefficients of $F_\Ua$
in the Lagrangian prior to the substitution~\myref{jjhe}.

On the other hand if we had treated $\Ua$ as an ordinary
chiral superfield of $U(1)_K$ weight 2, we would have
\beq
\Ds \Ua - \Dbs \Uba = t.d.,
\eeq
and therefore
\beq
F_\Ua - \Fbar_\Uba = t.d.
\eeq
The supersymmetry breaking vev disappears on the right-hand
side in this approach, leading to a different phenomenology.
Incidentally, this also occurs in the approach
taken in \cite{DQQ94},
where it was assumed that the highest component of the
Chern-Simons superfield $\Omega_\ap$ (see below) vanishes.  But this is
nothing other than $F_\Ua$, which as we have seen
does not have a vanishing vev in the BGW
approach if supersymmetry is broken.  In fact, if we
look at the perturbative form of $\Omega_\ap$ as given
in \cite{DQQ94}, there is no apparent reason why
its highest component would not get a vev, since
it is a scalar operator that solely contains
strongly interacting fields.

One might think that it would
be useful to relate the effective Lagrangian
to that of the chiral dilaton formulation and
compare in the global limit where the effective
theory is known \cite{ADS}.
However, the last term on the right-hand side
of \myref{jjhe} is implicitly suppressed by powers
of $1/m_P$, the inverse reduced Planck mass,
and is obviously a supergravity
effect.  (After all, it involves the auxiliary
scalar $M$ of supergravity.)  Hence it would not
appear the theories of global supersymmetry.
Nevertheless it is important in the present
context because the soft term phenomenology
will generally be affected by its presence.

In conventional supergravity coupled to super-Yang-Mills
and a chiral dilaton,
the chiral field strength superfields $\W_\ap^\alpha$ are introduced
through an {\it F-density,} which necessitates a holomorphic
metric $f_{\ap (b)}(S,\Phi)$; i.e., on has
\beq
\Lcal_{\stxt{YM}} = \sint {E \over 8R} f_{\ap (b)}(S,\Phi)
(\W^\ap \W^{(b)}) + \hc
\label{jjyt}
\eeq
However, it is well-known that this is not the unique
local superfield Lagrangian through which the Yang-Mills
field strength can be introduced; the possibility of
a nonholomorphic metric is allowed if instead we introduce
the chiral field strength superfields using
a {\it D-density} Lagrangian \cite{CFGVP}.  Generically,
the linear multiplet approach leads to a super-Yang-Mills
Lagrangian that is equivalent to a combination of the
holomorphic F-density and nonholomorphic D-density.
On the other hand it has been shown that
the one-loop effective Lagrangian derived from
heterotic orbifold models \cite{DKL91} is such that one can
always write the super-Yang-Mills
Lagrangian as a pure F-density \cite{DFKZ92}.
In the linear dilaton formulation matched to
the string theoretic calculations, one of course
obtains a super-Yang-Mills Lagrangian that is just an
F-density in the chiral dilaton
formulation.\footnote{The component expansions given
below are general enough to accomodate either situation;
to obtain only an F-density
requires that arbitrary functionals that
appear here satisfy certain conditions.}
But, it would be interesting to formulate the
corresponding conditions explicitly in the
general context.  The component expansion provided
here may be of some aid in such an enterprise.

The advantages of the linear multiplet listed
above suggest that
in detailed model building---which intends to go beyond
leading order, implement gaugino condensation,
and nonperturbative corrections to the dilaton
K\"ahler potential---the linear dilaton
is a practical tool.  At the very least, it is worthwhile
to have parallel studies in a different formulation
which is supposed to be equivalent to the chiral dilaton.
These are among the reasons for which
BGW chose to work in this setting.
It then becomes useful to have component field
expansions that are general enough to handle the
cases envisioned in semi-realistic applications.  This is
the motivation for the computation reported here.

To properly relate the linear dilaton
to the chiral dilaton,
the duality transformation should respect
supersymmetry.  One approach is to perform
the duality transformation at the superfield
level \cite{Sie79}.
For global supersymmetry, this duality has been reviewed
in Sections 2 of \cite{DQQ94} and \cite{GG99}.  For the
locally supersymmetric (supergravity) case
in the $U(1)_K$ formalism, a review of
chiral-linear duality has been given (briefly)
in BGG Section 5.5.  The duality in the
superconformal approach has been discussed in
\cite{DFKZ92,DQQ94}.

For illustrative purposes, consider the Lagrangian
\beq
\Lcal = \sint E \[ -2 + f(L) + \third (L + \Omega) (S + \bar S) \] .
\label{etae}
\eeq
Here $L$ is treated as a real superfield which is otherwise
unconstrained; that is, $\Lcal$ is supposed to represent
a {\it first-order formulation} of the {\it target theory.}
$S$ is a chiral superfield.  $\Omega$ is real and is
the {\it Chern-Simons superfield} \cite{CFV87}
(see also \cite{DFKZ92} or Appendix F.3 of BGG).
It satisfies the constraints
\beq
\chiproj \Omega = \tWW, \qquad \chiprojb \Omega = \tWWb .
\eeq
The superfield equations of motion obtained from
\myref{etae} yield the duality
\beq
{L \over 1 + f(L)} = {1 \over S + \bar S}
\label{yath}
\eeq
together with the modified linearity conditions \myref{plpj}.
We denote the \ttz\ ({\it lowest}) components as
\beq
L\| = \ell, \qquad S\| = s, \qquad \bar S\| = \bar s .
\eeq
Above we mentioned that in the chiral formulation
the string coupling $g_s$ is determined by the real part of $s$;
the duality relation \myref{yath} gives a corresponding
meaning to $\ell$:
\beq
g_s^2 = {2 \over s + \bar s} = {2 \ell \over 1 + f(\ell)} .
\label{ahqq}
\eeq
Further details on the duality relations---for
the case of local supersymmetry in the
$U(1)_K$ formalism, including
the other component fields---may be found in
\cite{BGGMb,BGGa}.

A comparison of phenomenological
implications of either formulation is given
in~\cite{BGW97a}.  In this work it was
found that the K\"ahler moduli of the underlying
theory are not stabilized at their {\it self-dual
points} in the chiral dilaton approach, whereas
the stabilization does occur at the self-dual
points in the linear dilaton approach.  It was argued
that the disparate results originate from
an explicit $S$ dependence in the effective
superpotential of the chiral dilaton formulation.

However, if one starts with a linear dilaton
and performs a duality transformation analogous
to \myref{etae}, it {\it must} be that one obtains
a chiral dilaton formulation which is equivalent
{\it on-shell.}  That is, the first-order formalism
{\it ensures} that the equations of motion for the
linear dilaton supergravity are equivalent
to the equations of motion for the chiral dilaton
supergravity, and that the equivalence is
established through the superfield redefinition
that is obtained through the duality transformation---the
generalization of \myref{yath}.  The stabilization
of moduli is studied through minimization of the
scalar potential.  But this is nothing but a
study of solutions to the equations of motion
in the infrared limit, neglecting all
fields with nonzero spin.  Since the equations are
equivalent in the two approaches, they must yield
equivalent solutions.
Thus it is our opinion that some subtlety must
have been overlooked in performing the
duality tranformation for the
theory studied in \cite{BGW97a}.
In that case the target theory was more
complicated than the illustrative example
\myref{etae}.  We intend to return to this
issue in a future publication.
A full component
lagrangian may shed some light on this issue,
since it allows us to study the duality
transformations at the component field level.

\subsection{Antecedants}
The effective supergravity discussed here is
an extension of the BGW effective theory \cite{BGW96,BGW97a,BGW97b},
which does not include an anomalous \uone\ factor (hereafter
denoted \ux) in the gauge group.
A \ux\ is a generic feature of semi-realistic string
constructions; for example, in \cite{Gie02a} it
was found that 168 of 175 models had a \ux.
The associated anomaly is cancelled by a
{\it Green-Schwarz (GS) counterterm} \cite{GS84,UXR},
as will
be discussed below.
The linear multiplet formulation provides an
elegant description of the effective supergravity
that results, as has been discussed in \cite{GG02a}.
Indeed Refs.~\cite{GG02a,GG02b} as well as work in
progress \cite{GG02c} aim to address the modifications
to the BGW effective theory
in the presence of an anomalous \uone.  However,
in none of these references is
the full fermionic Lagrangian presented; only
the gravitino and gaugino effective masses have been
computed \cite{BGW97a,BGW97b}.  Moreover,
we allow for unconfined matter to couple to the
(auxiliary) hidden matter
condensate superfields $\Pi^\alpha$.  This is important for
the stabilization of flat direcions in the
presence of an anomalous $U(1)$ factor,
so-called {\it D-moduli} \cite{DMod}.

Fermion terms of component Lagrangians in the
linear dilaton formulation
have previously been computed by authors
other than BGW to varying degrees.

In \cite{ABGG93}, Adamietz et al.~obtained {\it all}
the fermionic terms.  However, no superpotential was included
in the Lagrangian, a GS counterterm
for a \ux\ was not included, and an effective theory of gaugino
condensation was not explicitly added.
Adamietz et al.~also made the simplifying
assumption that the K\"ahler potential for the
linear multiplet is $k(L) = \alpha \ln L$,
which is equivalent to the assumption
$K(S,\bbar{S}) \propto - \ln(S + \bbar{S})$
in the chiral dilaton approach.
Stabilization of the dilaton sometimes
requires a more general function,
such as will be studied here.

In~\cite{DQQ94}, Derendinger et al.~only gave
some of the fermion terms; in particular gaugino bilinears.
Their treatment of gaugino condensation differs
from that of BGW in some important ways, as
will be discussed below; these differences
affect predictions for soft supersymmetry
breaking operators in the low energy effective
theory.  Also, a GS counterterm for a \ux\ was not included
in the effective theory.
Derendinger et al.~use the superconformal tensor
calculus \cite{KU83} to obtain the component Lagrangian,
whereas we use $U(1)_K$ superspace.  We believe that
it is useful to have results in both formalisms.

Various other limiting assumptions were made in
these previous works which have not been made here.
Thus our calculation can accomodate a more general
set of circumstances and exhibits possible couplings
that were not accounted for in previous works.

\subsection{The Lagrangian}
In this article a very general K\"ahler
potential is assumed; it is the same as
for BGW~\cite{BGW96,BGW97a,BGW97b}:
\beq
K = k(L) + G(\Phi,\Phibar), \qquad k(L) = \ln L + g(L).
\label{hahr}
\eeq
Here $g(L)$ is left arbitrary in our calculations,
though we have in mind the sort of nonperturbative
corrections that are expected based on general
arguments \cite{She91} and string duality \cite{Sil97}.
Indeed these sorts of corrections have been used
by BGW and others to stabilize the dilaton at weak coupling
(i.e., $g_s^2 \lappeq 1$ in \myref{ahqq})
in a scheme that has come to be known as {\it K\"ahler
stabilization} \cite{BGW96,BGW97a,BGW97b,Cas96,Wu96,Bar98,KLM01}.

The Lagrangian consists of several pieces:\footnote{See
\cite{BGW97a} for
further discussion of the significance of each term.}
\beq
\Lcal = \Lcal_{\stxt{kin}} + \Lcal_{\stxt{pot}} +
\Lcal_{\stxt{VY}} + \Lcal_{\stxt{thr}}+ \Lcal_{\stxt{GS}}^0
+ \Lcal_{\stxt{GS}}^X .
\eeq
The first piece contains the usual kinetic terms for
all the fields, and is written in the $U(1)_K$
superspace formalism as follows:
\beq
\Lcal_{\stxt{kin}} = \sint E \[ -2 + f(L) \] .
\label{rar1}
\eeq
The function $f(L)$ is chosen such that a canonical Einstein
term $-\half {\cal R}$ (where ${\cal R}$ is the
Ricci scalar) is obtained in the component expansion.
With reference to the K\"ahler potential \myref{hahr},
the condition for this to be true is that
\beq
L g'(L) = f(L) - L f'(L),
\eeq
where $g'(L) = dg(L)/dL$, etc.  An elementary discussion of
how this condition occurs can be found
in Section 5.4 of BGG, where their
function $F$ is related to the $f$ used here according
to $F=(2-f)/3$.

The usual superpotential term is included:
\beq
\Lcal_{\stxt{pot}} = \sint {E \over 2R} e^{K/2} W(\Phi,\Pi) + \hc
\label{kjsr}
\eeq
We remind the reader that
the chiral superfields $\Pi^\alpha$ are static fields
corresponding to matter condensates of the hidden sector.
Thus they do not appear in \myref{hahr}, but it is
important to include them in \myref{kjsr}.

In addition to this {\it bare} superpotential, we
have the Veneziano-Yankielowicz effective superpotential \cite{VY82},
with suitable modifications suggested by Taylor \cite{Tay85}:
\beqa
\Lcal_{\stxt{VY}} &=& \sint {E \over 8R} \sumC U_\ap
\[ {\over} b_\ap' \ln \( e^{-K/2} U_\ap / \mu^3 \)
\right. \ddd \left.
+ \sum_\alpha b_\ap^\alpha \ln
\( A_\ap^\alpha(\Phi) \, \Pi^\alpha \) \] + \hc
\eeqa
The gaugino condensate superfields $U_\ap$ appear
explicitly here.  When they are integrated out we
obtain the usual nonperturbative superpotential \cite{ADS}
induced by instanton effects,\footnote{Indeed,
one can argue that $\Ua$ have
masses of order the condensation scale and {\it should}
be integrated out to obtain the effective theory
below that scale.  We thank
Erich Poppitz for a remark in this regard.} only coupled to supergravity
in the present context; to match to the globally
supersymmetric results one should take the decoupling
limit $m_P \to \infty$.  The coefficients
$b_\ap'$ and $b_\ap^\alpha$ are constrained by a matching
to the weak coupling quantum anomalies.  A further
discussion can be found, for example, in \cite{BGW97a}.

Massive string states
can yield threshold corrections to the effective
theory below the string scale.  The well-known
corrections associated with $N=2$ sectors in
orbifold compactifications of the heterotic
string are given by \cite{DKL91,ANT91}:
\beqa
\Lcal_{\stxt{thr}} &=& \sum_I \sint {E \over 8R}
\[ \sumNC b_\ap^I(\W \W)_\ap
\right. \ddd \left.
+ \sumC b_\ap^IU_\ap \]  \ln \eta^{-2}(T^I) + \hc
\label{uyw}
\eeqa
where the coefficients $b_\ap^I$ are determined
by explicit string calculations.

Quantum anomalies that arise from the
terms so far described are cancelled by Green-Schwarz (GS)
counterterms.
The first involves a real function $S$---not to be
confused with the chiral dilaton of the discussion above---which
we will refer to as the {\it GS potential.}
We restrict $S$ to be a function of chiral superfields:
$S = S(\Phi,\Phibar)$.  Its purpose is to
cancel target-space duality anomalies.
Its superfield expression is:
\beq
\Lcal_{\stxt{GS}}^0 = \sint E L S.
\label{rar2}
\eeq
The precise form of $S$ can only be obtained from
a detailed understanding of the full anomaly structure
of the effective supergravity and how it is canceled
in the underlying string theory.  The
{\it modular anomaly} associated with $SL(2,Z)^3$
tranformations on {\it K\"ahler moduli} associated
with the complex planes in orbifold compactifications
of the heterotic string is well-known.  It is
partially canceled by, for example, a choice of $S \propto G$,
where $G$ is identified in \myref{hahr}.  However,
a richer anomaly structure is anticipated on the
basis of 1-loop supergravity calculations \cite{PVREG,anom},
and so we leave $S$ arbitrary in our component expansion.
We note that since
the GS counterterm potential $S$ is left in a
rather general form, our component field
expansions do not assume modular invariance;
that is, we can accomodate models where violations
of modular invariance are envisioned, due
to nonperturbative effects in the underlying
string theory.  On the other hand, exact
modular invariance can also be imposed with
an appropriate choice for $S$.

The second GS counterterm is associated with the
anomalous \ux, with a corresponding
vector superfield $V_X$.  It is given by:
\beq
\Lcal_{\stxt{GS}}^X = \sint E L \delta_X V_X .
\label{lljr}
\eeq
This addition to the BGW effective theory has
been the subject of recent work \cite{GG02a,GG02b,GG02c}.
There it was shown how to fix to unitary gauge
and integrate out the modes that acquire
large masses when the Fayet-Iliopoulos (FI) term that arises from
\myref{lljr} spontaneously breaks \ux\ at a high scale.

We have omitted the perturbative one-loop effective quantum
correction to the Lagrangian.  Related expresssions have been studied
by various groups:  by Derendinger et al.~using superconformal
methods~\cite{DFKZ92}; by Bagger et al.~using
a component field approach \cite{Bagger:1999rd};
by Gaillard et al.~in $U(1)_K$
superspace~\cite{GNW}.  However, all of these calculations
involve various simplifying assumptions on the
form of the bare Lagrangian compared to what
is given here.  Furthermore, in the calculation of
\cite{GNW} there exist some uncertainties
in the precise form of the chiral projection operator $P_\chi$
employed there.  In principle the one-loop effective quantum
correction to the Lagrangian can be derived from
$\Lcal_{\stxt{kin}} + \Lcal_{\stxt{pot}}$
by a one-loop computation using, say, Pauli-Villars regularization
\cite{PVREG}.
In fact, much of this calculation has been performed
in \cite{PVREG} for the class of Lagrangians studied here.
One possible motivation for the present work
is to fill in the fermionic details of the component Lagrangian
needed to complete the one-loop computation.

\mysection{Aspects of the component Lagrangian}
\label{resu}
Our results for
$\Lcal_{\stxt{pot}} + \Lcal_{\stxt{VY}} + \Lcal_{\stxt{thr}} + \Lcal_{\stxt{GS}}^X$
are given in Appendix \ref{ezpt}.
For this part of $\Lcal$ the component field expansion is straightforward,
except for a certain subtlety that arises in
$\Lcal_{\stxt{GS}}^X$.  This has to do with the
evaluation of spinorial derivatives acting on the
$U(1)_X$ vector superfield $V_X$.  Here it is important
to properly account for the conventions of BGG for
the solution of superspace Bianchi identities;
details are given in Appendix \ref{cna}.

The superpotential Lagrangian \myref{spcf} contains
the usual terms that are present in chiral supergravity; of course
a mixing with the dilaton occurs due to the $\ell$
dependence in the $e^{K/2}$ prefactor for these
terms.\footnote{We remind the reader of the dilaton ($\ell$) dependent
contribution $k(\ell)$ to the K\"ahler potential \myref{hahr}; also
note that $k'=\p k/\p \ell$, etc.~below.}  In addition we have
pieces explicitly associated with the linear supermultiplet:
\beqa
\ove \Lcal_{\stxt{pot}}^L &=&
e^{K/2} \left\{ - \fourth W(k'' + k'^2) \vps
- {1\over \sqtw} \( W_k + W G_k \) k' (\vp \chi^k)
\right. \ddd \left.
+ W k' \[ \fourth \ub - \fourth \tr (\lb \lb) + \third \Mb \ell
+ {i \over 2} (\psibar^m \sbar_m \vp) \] \right\} + \hc
\label{scf1}
\eeqa
The bosonic terms were previously studied in the works of BGW.
Note that the (effective) dilatino and gaugino
masses receive contributions from the bilinears
$\vp \vp$ and $\lb \lb$ that appear in \myref{scf1}.
The bilinear $\vp \chi^k$ which mixes the dilatino $\vp$ with
matter fermions $\chik$ is a feature that
deserves further study.\footnote{The coupling of the
dilatino to the gravitino can be eliminated with the
``gauge'' choice $(\psibar^m \sbar_m)^\alpha = 0$.}
In particular, the presence of large vevs generally leads to
important effects that would arise from the $\vp \chi^k$
bilinear.  For example, it is common in semi-realistic
string models for exotic states to be removed at a
high scale through large effective masses generated by
FI-induced vevs:
\beq
m_{ij} \sim \half \vev{e^{K/2} W_{ij}} \lappeq \ord{0.1} m_P,
\label{jar3}
\eeq
where $m_P$ is the reduced Planck mass.
This implies effective couplings in \myref{scf1} of the form
\beq
- \sqtw {m_{ij} \over m_P} k'(\ell) \phi^i (\vp \chi^j).
\eeq
The implications of such couplings for the cosmology associated with
the dilaton, dilatino and heavy matter states presents an interesting
topic for further study.

A feature that is special to
the fermionic terms appears from the effective theory
of gaugino condensation.  This is related to the auxiliary
fermions $\La_\ap^\alpha$ contained in the
gaugino condensate superfields $\Ua$;
see \myref{krrt}.  When these fields are eliminated by
their equations of motion, we obtain their contribution
to the Lagrangian:
\beqa
\ove \Lcal(\La) &=& \sumC {b'_\ap \over 8 \ua} (\La \La)_\ap + \hc
\nnn &=& \sumC {2 \ua \over b'_\ap} \left\{ (f^\ap)^2 \vps
+ \hat f_k^\ap \hat f_\ell^\ap \ckl
\right. \ddd
+ (\tilde f^\ap)^2
(\psib_m \sbar^m \s^n \psib_n) + 2 f^\ap \hat f_k^\ap
(\vp \chik)
\ddd \left. + 2i \tilde f^\ap \[ f^\ap (\psib_m \sbar^m \vp)
+ \hat f_k^\ap (\psib_m \sbar^m \chik) \] \right\} + \hc ,
\label{qwqq}
\\
f^\ap &=& {1 \over 4 \sqtw} \[ b'_\ap k' - f'' - \(
{k' + \ell k'' \over 1 - \third \ell k' } \) k'
- 2 k''
\right. \ddd \left.
- \( { k' + 3 k'' \ell \over k' \ell - 3} \) k' \] ,
\\
\hat f_k^\ap &=& \fourth \( b'_\ap G_k - h_k^\ap - S_k \) ,
\\
\tilde f^\ap &=& {1 \over 8 \sqtw} \[ 2 b'_\ap
\ln \( e^{1 - {K \over 2}} \ua / \mu^3 \) + 2 h^\ap
+ f' + k' + S \] .
\eeqa
The (holomorphic) function $h^\ap(\phi,\pi)$ is defined
in \myref{qwqw}, and
\beq
h_k^\ap \chik = {\p h^\ap \over \p \phik} \chik
+ {\p h^\ap \over \p \pi^\alpha} \chi^\alpha .
\eeq
Here $\sqtw \chi^{\alpha \beta} = \D^\beta \Pi^\alpha \|$ is
a further auxiliary spinor, contained in the
matter condensate superfield, which can likewise
be eliminated using its equations of motion.

As can be seen from \myref{qwqq},
we obtain a {\it soft mass} for the
dilatino; it is roughly the order of the supersymmetry
breaking scale $\ua/m_P^2 \sim 1$ TeV.
Furthermore, matter fermions get a soft mass
contribution, which includes
mixing with the dilatino,
suppressed by $\vev{\hat f_k^\ap}/m_P$.
These must be singlets under the {\it Standard Model}
gauge group $SU(3)_C \times SU(2)_L
\times U(1)_Y$ in order for this effect to matter.
To see this note that $\hat f_k^\ap$ transforms as the conjugate
of $\chik$. This implies $\vev{\hat f_k^\ap}/m_P
\lappeq 10^{-15}$ if $\chik$ is charged under
the Standard Model.  Thus these terms are relevant only
in extended models, such as the {\it Non-Minimal
Supersymmetric Standard Model (NMSSM),} or models
with an {\it inflatino,} etc.
Going to the gauge $\psib_m \sbar^m \equiv 0$ demonstrates
that the above terms do not contribute to the gravitino
mass.

In the language of \cite{DQQ94}, the auxiliary
spinors $\La_\ap$ and ${\bbar{\La}}_\ap$ correspond
to the next-to-highest components of the Chern-Simons
superfield $\Omega_\ap$.  However, the authors of
\cite{DQQ94} set these components identically to
zero (cf.~their Eq.~(3.29));
hence, the above contributions to soft fermion
masses do not appear in their results.
This is one of the ways in which our results
generalize previous work.

\mysection{Conclusions}
\label{conc}
In this note we have carried out
the calculation of the component Lagrangian
for supergravity coupled to gauged matter and a linear
dilaton multiplet.  We have reviewed the reasons
why this alternative to the chiral dilaton
formultion might be useful.  We have commented
on previous work that exists in the the literature
and have explained the ways in which the results
presented here generalize those that have
appeared before.  We have offered ways in
which these results might be put to use.

In particular we believe that further details
of the duality between the linear dilaton and
the chiral dilaton should be explored.  Since
some results in the literature are at odds
it would appear that a more careful comparison
at the component field level may uncover the
errors which we suppose have led to these
discrepancies.  At the same time, holomorphy
prevents corrections to the superpotential that
involve the linear multiplet and these must
arise as exact symmetries in the chiral formulation.
It would be interesting to make the connection
more precise.  The component expansion provided
here makes that possible since the duality transformations
can be checked at the component field level.

K\"ahler stabilization involves deviations of
the dilaton K\"ahler potential from the leading
order form.  Such corrections are easily
encoded with the modular invariant $L$, whereas
they require the modified dilaton multiplet $S'$
which mixes $S$ and $T^I$ in the chiral formulation.\footnote{That
is, one would typically introduce nonperturbative
corrections to the dilaton K\"ahler potential in
a duality invariant way; see
for example \cite{DFKZ92} for a definition of $S'$.}
This is no real impediment, but it does make the
two frameworks difficult to relate; we can imagine
that each {\it naturally} probes some different
regions of {\it parameter space} for these
nonperturbative corrections.  A further study of
the duality transformations is required to determine
the extent to which a representative coverage
is achieved in either scheme.

We have also pointed out some of the peculiarities
of the component Lagrangian.  The fact that $L$
cannot appear in the superpotential leads to special
constraints that yield a more restrictive phenomenology
if they are respected.  Whereas they can be
imposed in the chiral dilaton formulation, they
``fall out'' in the present work.  This is a nice
feature because in a certain sense it automates
model-building.  We have also pointed out how the
effective theory of dynamical supersymmetry breaking
is impacted by the fact that the gaugino condensate
superfield is obtained from $L$ through Bianchi
identies; in particular we showed how this can
impact the soft term phenomenology of the low
energy theory.

As can be seen, several details of the effective
theory, and its relation to other formulations,
remain to be explicitly sorted out.  We do not
expect any remarkable things to be found through
further exploration of this duality, but we
anticipate that complete agreement between the
formulations will emerge.  As this goal is achieved,
the related phenomenological studies will become
increasingly reliable and accurate.  Furthermore,
the situations that can be studied easily will be
enlarged by the availability of component Lagrangians
that are more general and can thus accomodate
a greater variety of assumptions at the superfield
level.

\vspace{15pt}

\noindent {\bf \large Acknowledgements}

\vspace{5pt}

\noindent
The author would like to thank Mary K.~Gaillard,
Brent D.~Nelson and Erich Poppitz for useful discussions.
Thanks are also owed to the referee, who
provided helpful suggestions for how this article might be
improved.  This work was supported in part by the
Director, Office of Science, Office of High Energy and Nuclear
Physics, Division of High Energy Physics of the U.S. Department of
Energy under Contract DE-AC03-76SF00098 and in part by the National
Science Foundation under grant PHY-00-98840.
Support was also provided by NSERC.

\myappendix

\mysection{Notation and conventions}
\label{cna}
The linear superfield $L$ is defined to satisfy modified
linearity conditions such that
\beqa
\chiproj L &=& -U - \tWW, \qquad \chiprojb L = - \Ub - \tWWb,
\label{cic} \\
{[} \Dad, \Dbadd ] L &=& 4 L \Gaad + 2 \Baad + 2 \tr (\W_\alpha \Wb_\adot),
\label{cid}
\eeqa
where we abbreviate sums over $G_C$ (condensing parts of the
gauge group) and non-$G_C$ parts of
the gauge group by
\beq
\tWW \equiv \sumNC \WWa, \qquad  U \equiv \sumC \Ua, \qquad {\rm etc.}
\label{sdf}
\eeq

In what follows we adopt the conventions and notation of BGG for the
definitions of component fields in terms of \ttz\ parts
(denoted by $\|$)
of spinorial derivatives ($\Dad, \Dbadu$, etc.)
of superfields, with the exception
that we denote the dilatino according to:
\beq
\vp_\alpha \equiv \Dad L \|, \qquad \vpb^\adot \equiv \Dbadu L \|.
\eeq
In addition we define the component fields
\beq
\ua = \Ua \|, \qquad
\La_{\ap \alpha} = {1 \over \sqtw} \Dad \Ua \|, \qquad
F_\Ua = -\fourth \Ds \Ua \|,
\label{krrt}
\eeq
and corresponding conjugates.  We also have in the
notation of \myref{sdf}
\beq
u = \sumC \ua, \qquad
\Lambda_\alpha = \sumC \Lambda_{\ap \alpha},
\qquad F_U = \sumC F_\Ua.
\eeq
For the (auxiliary) matter condensate superfields $\Pi^\alpha$
we have component fields
\beq
\pi^\alpha = \Pi^\alpha \|, \qquad
\chi^\alpha_\beta = {1 \over \sqtw} \D_\beta \Pi^\alpha \|, \qquad
F^\beta = - \fourth \Ds \Pi^\alpha \|,
\eeq
where $\alpha$ should not be confused with a spinor index.

A semicolon denotes the usual K\"ahler covariant
differentiation on the complex scalar manifold; e.g.,
\beq
W_{k;\ell} = W_{k\ell} - \Gamma_{k\ell}^m W_m,
\qquad \Gamma_{k\ell}^m = G^{m \bar m}
G_{k \bar m \ell}.
\label{juur}
\eeq
In addition to the usual gauge and space-time reparameterization
covariance, K\"ahler covariance and $U(1)_K$ covariance
is included in the covariant derivatives that appear
in the component expansions.
E.g., for the fermionic
components of chiral superfields we have
\beq
\D_m \chiak = \p_m \chiak - \omega_{m\alpha}^{\spc \spc \beta}
\chi_\beta^k - A_m \chiak - i a_m^\ap (T_\ap \chi_\alpha)^k
+ \chi^i \Gamma^k_{ij} \D_m \phi^j,
\eeq
and for the dilatino we have
\beq
\D_m \vpau = \p_m \vpau + \vp^\beta \omega_{m \beta}^{\spc \spc \alpha}
- \vpau A_m .
\label{cddf}
\eeq
Here, $\omega_{m\alpha}^{\spc \spc \beta}$ is the
usual spin connection, $a_m^\ap$ is the Yang-Mills connection,
$A_m$ is the $U(1)_K$ connection, and the K\"ahler
connection $\Gamma^k_{ij}$ is defined in \myref{juur}.
The coefficient of the $A_m$ term in $\D_m$ depends
on the $U(1)_K$ weight of the field on which
it acts.\footnote{See
Section 4 of BGG for a fuller specification and explanation of
covariant derivatives in the present formalism.}
The component field expansion for $A_m \equiv A_m \|$ is given by
\beqa
A_m \| &=& -\iotw b_m + \fourth G_k \D_m \pk
- \fourth G_\kbar \D_m \pbk
+ {i \over 4} G_{k \kbar} (\chi^k \s_m \chib^\kbar)
\ddd
- {i \over 8} k'' (\vpb \sbar_m \vp)
+ {i \over 6} k' \ell b_m - {i \over 4} k' B_m
\ddd
- {i \over 8} k' \tr (\lb \sbar_m \la)
+ \eig k' (\psi_m \vp) - \eig k' (\psib_m \vpb) .
\label{haq2}
\eeqa
We have
checked that our expression for $A_m \|$ is equivalent
to (BGG-E.3.4) in the special case of $k(L)=\alpha \ln L$;
in this calculation (BGG-5.2.20) and (BGG-5.3.7) are
especially useful; furthermore, a typo in (BGG-E.3.4)
must be corrected---the ``level'' factor of $k$ (not to be
confused with the functional $k(L)$) should be
absent on the ${}^*h_m$ term that appears in (BGG-E.3.4).

We evaluate the terms of $\Lcal_{\stxt{GS}}^X$ and its spinorial derivatives
in Wess-Zumino (WZ) gauge:
\beq
V_X \| \WZ \Dad V_X \| \WZ \Dbadd V_X \| \WZ \Dad \Dbd V_X \|
\WZ \Dbadd \Dbbdd V_X \| \WZ 0.
\label{wzd}
\eeq
To evaluate the component field expansions of the
spinorial derivatives of $V_X$,
we must be careful to use the conventions of BGG for the
solution to the superspace Bianchi identities,
and {\it not} those of, for example, Wess and Bagger \cite{WB92}.
Taking this into account, we find:
\beqa
\Dad \Dbadd V_X \| &\WZ& -a_{X \alpha \adot}, \nnn
\Dad \Dbs V_X \| &\WZ& 4i \la_{X \alpha} + 2i a_{X m}
(\s^n \sbar^m \psi_n)_\alpha ,
\nnn
\Ds \Dbs V_X \| &\WZ& 8 \Dbf_X + {16 \over 3} b^m a_{Xm}
- 4 a_{Xn} (\psi^m \s^n \psib_m) - 8i \D^m a_{Xm} \nnn
&& + 4 (\lb_X \sbar^m \psi_m) - 4 (\psib^m \sbar_m \la_X) .
\eeqa
These expressions are sufficient to compute $\Lcal_{\stxt{GS}}^X$,
when combined with other identities given here and in BGG.

It proves convenient to introduce the following
abbreviations:
\beqa
\Delta_m \phi^k &=& e_m^{\spc a} \D_a \Pk \|
= \D_m \phi^k - {1 \over \sqtw} (\psi_m \chi^k) ,
\label{cdph} \\
\Delta_m \phib^\kbar &=& e_m^{\spc a} \D_a \Pbk \|
= \D_m \phib^\kbar - {1 \over \sqtw} (\psib_m \chib^\kbar) ,
\label{cdpb} \\
\Delta_m \ell &=& e_m^{\spc a} \D_a L \|
= \p_m \ell - \half (\psi_m \vp) - \half (\psib_m \vpb) ,
\label{cddi} \\
i \hf_{\ap nm} &=& if_{\ap nm} + (\psi_n \s_m \lb_\ap)
+ (\psibar_n \sbar_m \la_\ap) .
\label{daz2}
\eeqa
Because of its length, we find it convenient to
abbreviate $\Ds \WWa \|$ below.
It is straightforward to obtain $\Ds \WWa \|$ from (BGG-4.5.25)
if one makes the identification $f_{(r)(s)} \equiv -16$ for
the functional $f_{(r)(s)}$ that appears there:
\beqa
\lefteqn{\Ds \WWa \| =}
\ddd
2 f_\ap^{mn} f_{\ap mn} + i \e^{mnpq} f_{\ap mn} f_{\ap pq}
+ 8i (\la \s^m \D_m \lb)_\ap
\ddd
- 4 \Dbf_\ap^2
+ 4 \Mb (\la \la)_\ap
- 4 (\la_\ap \s^m \psib_m) \Dbf_\ap
\ddd
- 4i \[ (\psi_m \s^{pq} \s^m \lb_\ap)
+ (\psib_m \sbar^{pq} \sbar^m \la_\ap)
- (\psib_m \sbar^m \s^{pq} \la_\ap) \] f_{\ap pq}
\ddd
- 2 \[ (\psi_m \s^{pq} \s^m \lb_\ap)
+ 2 (\psib_m \sbar^{pq} \sbar^m \la_\ap)
- (\psib_m \sbar^m \s^{pq} \la_\ap) \]
\ddd
\qquad \times \[ (\psi_p \s_q \lb_\ap) + (\psib_p \sbar_q \la_\ap) \] .
\label{dwx}
\eeqa
It is also useful to abbreviate the following
components of the superspace torsion:
\beqa
T_{cb}^{\spc \alpha} \| &=&
\half e_b^{\spc m} e_c^{\spc n} \( \D_n \psi_m^\alpha
- \D_m \psi_n^\alpha \)
\ddd
+ {i \over 12} \[ e_c^{\spc m} (\psi_m \s_n \sbar_b)^\alpha
- e_b^{\spc m} (\psi_m \s_n \sbar_c)^\alpha \] b^n
\ddd
- {i \over 12} \[ e_c^{\spc m} (\psib_m \sbar_b)^\alpha
- e_b^{\spc m} (\psib_m \sbar_c)^\alpha \] M,
\label{tor1} \\
T_{cb \adot} \| &=&
\half e_b^{\spc m} e_c^{\spc n} \( \D_n \psib_{m \adot}
- \D_m \psib_{n \adot} \)
\ddd
- {i \over 12} \[ e_c^{\spc m} (\psib_m \sbar_n \s_b)_\adot
- e_b^{\spc m} (\psib_m \sbar_n \s_c)_\adot \] b^n
\ddd
- {i \over 12} \[ e_c^{\spc m} (\psi_m \s_b)_\adot
- e_b^{\spc m} (\psi_m \s_c)_\adot \] \Mbar,
\label{tor2}
\eeqa
as can be found in (BGG-4.1.31) and (BGG-4.1.32).


\mysection{Projection to component fields}
\label{proj}
If $\Omega$ is a real superfield of $U(1)_K$ weight zero, then
we may use (BGG-D.1.10) to integrate by parts in superspace
and obtain
\beqa
&& \Lcal_\Omega \equiv \sint E \Omega  =  \sint {E \over 2R} \hat r_\Omega + \hc,
\ddd
\qquad {\rm where} \qquad
\hat r_\Omega  \equiv  -{1 \over 8} (\Db^2 - 8R) \Omega .
\label{ibpt}
\eeqa
Note that $\hat r_\Omega$ is the {\it chiral projection} of $\Omega$.
We use this technique to convert the integrals of \myref{uyw}
to the form \myref{ibpt}.  Doing so we have:
\beqa
\hat r_{\stxt{GS}} &=& \hat r_{\stxt{GS}}^0 + \hat r_{\stxt{GS}}^X,
\qquad
\hat r_{\stxt{thr}}
= \hat r_{\stxt{thr}}^{\stxt{P}} + \hat r_{\stxt{thr}}^{\stxt{NP}},
\nnn
\hat r_{\stxt{VY}}
&=& \hat r_{\stxt{VY}}^{U} + \hat r_{\stxt{VY}}^{\Pi} ;
\label{cab} \\
\hat r_{\stxt{pot}} &=& e^{K/2} W, \qquad
\hat r_{\stxt{GS}}^X  =  -{\dx \over 8} (\Db^2 - 8R) \( L V_X \) , \nnn
\hat r_{\stxt{thr}}^{\stxt{P}} & = &
{1 \over 4} \sum_I \sumNC (\W \W)_\ap
b_\ap^I \ln \eta^{-2}(T^I), \nnn
\hat r_{\stxt{thr}}^{\stxt{NP}}  &=&
{1 \over 4} \sum_I \sumC U_\ap
b_\ap^I \ln \eta^{-2}(T^I) \nnn
\hat r_{\stxt{VY}}^{U} & = & {1 \over 4} \sumC b_\ap' U_\ap
\ln \( e^{-K/2} U_\ap / \mu^3 \), \nnn
\hat r_{\stxt{VY}}^{\Pi}  &=&  {1 \over 4} \sumC
\sum_\alpha b_\ap^\alpha U_\ap \ln \( A_\ap^\alpha(\Phi) \, \Pi^\alpha \) ;
\label{caa} \\
\hat r_{\stxt{kin}} &=& -{1 \over 8} (\Db^2 - 8R) \[ -2 + f(L) \], \nnn
\hat r_{\stxt{GS}}^0  &=&  -{1 \over 8} (\Db^2 - 8R) \[ L S(\Phi,\Phibar) \] .
\label{cac}
\eeqa

It is convenient to introduce two holomorphic functionals
$\ha(\Phi,\Pi)$ and $\hha(\Phi)$ by the identifications
\beq
\hat r_{\stxt{VY}}^{\Pi}+\hat r_{\stxt{thr}}^{\stxt{NP}}
\equiv \fourth \sumC \Ua \ha(\Phi,\Pi), \qquad
\hat r_{\stxt{thr}}^{\stxt{P}}
\equiv \fourth \sumNC \WWa \hha(\Phi,\Pi) .
\label{iiop}
\eeq
From the expressions \myref{caa} we see that
\beq
\ha = \sum_\alpha b_\ap^\alpha \ln \( A_\ap^\alpha(\Phi) \, \Pi^\alpha \)
+ \hha, \qquad
\hha = \sum_I b_\ap^I \ln \eta^{-2}(T^I) .
\label{qwqw}
\eeq
It is worth noting that since our component expansions are
written in terms of $\ha,\hha$, our results are more
general than \myref{qwqw}, and accomodate any assumptions
of the form \myref{iiop}.

For any of the $\hat r_i$ defined above,
we have from (BGG-4.4.22) that
the corresponding component Lagrangian
is given in terms of the \ttz\ limit of
spinorial derivatives:
\beqa
\Lcal_i &=& e \[ - \fourth \D^2 \hat r_i \|
+ {i \over 2} \( \psibar_m \sbar^m \)^\alpha \D_\alpha \hat r_i \|
\right. \ddd \left.
- \( \Mbar + \psibar_m \sbar^{mn} \psibar_n \) \hat r_i \|
{\over} \] + \hc
\eeqa
Here, $e$ is the determinant of the ordinary vierbein;
i.e., the usual $\sqrt{-\det g}$ factor.
In each case $\hat r_i$ has $U(1)_K$ weight 2.  Let
the symbol $\D$ denote covariant differentiation
including $U(1)_K$ and $D$ covariant differentiation
not including $U(1)_K$.  Then
\beq
\D_\alpha \hat r_i = (D_\alpha + 2 A_\alpha) \hat r_i, \qquad
\D^2 \hat r_i =
\( D^\alpha + A^\alpha \) \D_\alpha \hat r_i,
\eeq
where the superform $A$ is the $U(1)_K$ connection.\footnote{The
quantity $A_m$ used above is related via $A_m = (E_m^{\spc B} A_B) \|$,
where $B$ runs over $b,\beta,\bdot$.}

\mysection{Geometric relations}
\label{geid}
Here we briefly discuss methods based on the
$U(1)_K$ superspace geometry that are
used in the more difficult expansions.
The first set arises in the kinetic part of the
Lagrangian $\Lcal_{\stxt{kin}}$, defined in \myref{rar1}.
The second set occurs in the Green-Schwarz counterterm Lagrangian
$\Lcal_{\stxt{GS}}^0$, associated with duality group invariance,
defined in \myref{rar2}.


The difficulty that
is encountered in evaluating $\Lcal_{\stxt{kin}}$
is the computation of the component field expansion of
\beq
\Dad R \|, \qquad \Dbadu \bar R \|, \qquad \Ds R \| + \hc
\eeq
Techniques for the evaluation of these in the presence of
a linear multiplet were developed in \cite{ABGG93}.
However in that case the $L$-dependent K\"ahler potential
is $k(L) = \alpha \ln L$ and some simplifications
occur; furthermore, our conventions for the
component field definitions differ slightly;
thus in the present context we must recalculate
these expansions.  We now detail
the techniques and arrange the
expressions that are evaluated in Appendices \ref{kine}
and \ref{mgse}.

For the evaluation of $\Dad R\|$, we appeal to the identity
(BGG-3.4.42):
\beq
\Dad R = -\third X_\alpha - \tthird (\s^{cb}\e)_{\alpha \gamma}
T_{cb}^{\bspc \gamma} ,
\label{rid}
\eeq
where $X_\alpha$ is the chiral field strength associated
with the K\"ahler potential:
\beq
X_\alpha \equiv -\eig \chiproj \Dad K.
\eeq
As originally described in \cite{ABGG93}, the field
strength $X_\alpha$ contains
$\Dad R$ because of the $L$-de\-pen\-dence in $K$ (cf.~Eq.~\myref{hahr})
and the modified linearity conditions \myref{cic}.
Thus we extract the $\Dad R$ contained
in $X_\alpha$ so that we can solve for it explicitly, following
the methods of \cite{ABGG93}---reviewed in BGG Section 5.4.
This involves the definitions:\footnote{
Note that the quantity $Y_\alpha$ appearing in (BGG-5.4.5)
is related to the present notation by $Y_\alpha \equiv
X_{0\alpha} + Z_\alpha$.}
\beqa
X_\alpha &=& X_{0\alpha} + Z_\alpha - L k'(L) \Dad R,
\nnn
X_{0\alpha} &\equiv& -\eig \chiproj \Dad G(\Phi,\Phibar),
\nnn
Z_\alpha &\equiv& L k'(L) \Dad R
- \eig \chiproj \Dad k(L) .
\label{opi}
\eeqa
Using \myref{opi} we rewrite \myref{rid} as
\beq
(k'L-3)\Dad R = X_{0\alpha} + Z_\alpha
+ 2 (\s^{cb}\e)_{\alpha \gamma} T_{cb}^{\bspc \gamma},
\label{oue}
\eeq
where now, as it turns out,
$\Dad R$ will not appear on the right-hand side
when we work out the component expansion.  Eq.~\myref{oue} is in
agreement with (BGG-5.4.6).
After considerable manipulation
we find for $\Dad R \|$ the
expansion given in \myref{zaex} and \myref{xzex}.
One obtains $\Dbadu \Rbar \|$ by hermitian conjugation.

For the evaluation of $(\Ds R + \hc)\|$ we appeal to
(BGG-3.4.44):
\beq
\Ds R + \hc = - {2 \over 3}R_{ba}^{\bspc ba}
- \third (\Dau X_\alpha + \hc) + 4 G^a G_a + 32 R \Rb .
\eeq
Similar to the situation described in
the previous paragraph, we need to extract
$(\Ds R + \hc)$ from $(\Dau X_\alpha + \hc)$ due to the
$L$-dependence in $K$.  With definition \myref{opi}
it is not hard to show that
\beqa
\lefteqn{(k' L - 3) (\Ds R + \hc) =}
\ddd
2 R_{ba}^{\bspc ba}
- 12 G^a G_a -96 R \Rb
+ (\Dau X_{0\alpha} + \Dau Z_\alpha + \hc)
\ddd
- (k' + k'' L) (\Dau L \Dad R + \hc) ,
\eeqa
in agreement with (BGG-5.4.8).  Taking the \ttz\
part of this expression yields
\beqa
\lefteqn{(k' \ell - 3) (\Ds R + \hc) \| =}
\ddd
 2 {\cal R}
- {4 \over 3} b^m b_m - {8 \over 3} M \Mb
+ (\Dau X_{0\alpha} \| + \Dau Z_\alpha \| + \hc) \nnn
&& - (k' + k'' \ell) (\vp^\alpha \Dad R \| + \hc) ,
\eeqa
where ${\cal R}$ is the space-time Ricci scalar.
Note that in the special case $k(\ell)=\alpha \ln \ell$ considered
in \cite{ABGG93} we have $k' + k'' \ell = 0$.  This
eliminates the last term and simplifies many of the
expressions given below.

The component field expansion of
$(\Dau X_{0\alpha}\| + \hc)$ is obtained from (BGG-4.2.13)
provided we take $K(\Phi,\Phibar) \to G(\Phi,\Phibar)$ in
their expressions for derivatives of the K\"ahler
potential.  This leaves only $(\Dau Z_\alpha \| + \hc)$ to be determined;
we have evaluated this in \myref{rar4}.

For the evaluation of $\Lcal_{\stxt{GS}}^0$ it was shown in
\cite{ABGG93} how to proceed through the chiral field
strength for the Green-Schwarz counterterm potential $S(\Phi,\Phibar)$:
\beq
X_{S \alpha} = -\eig \chiproj \Dad S .
\eeq
The evaluation of $X_{S \alpha}$ is more complicated than
that for $X_{0\alpha} \equiv -\eig \chiproj \Dad G$:
whereas $G_{k \kbar;\ell} \equiv 0$,
the corresponding quantity $S_{k \kbar;\ell}$ does
not necessarily vanish.  However, the organization
of the calculation around this field strength
proves productive and leads directly to the
results of Appendix \ref{mgse}.

\mysection{Expansion of
$\Lcal_{\stxt{pot}} + \Lcal_{\stxt{VY}} + \Lcal_{\stxt{thr}} + \Lcal_{\stxt{GS}}^X$}
\label{ezpt}
Here the expansions are straightforward.  See Appendix \ref{proj}
for superfield definitions of various parts of the Lagrangian
given here.
\beqa
\ove \Lcal_{\stxt{pot}} &=&
e^{K/2} \left\{ -\half \( W_{k ; \ell} + W G_{k ; \ell}
+ W_\ell G_k + W_k G_\ell + W G_k G_\ell \) \ckl \right.
\ddd  - \fourth W(k'' + k'^2) \vps
- \[ \Mbar + ( \psibar_m \sbar^{mn} \psibar_n ) \] W
\ddd  + \( W_k + W G_k \) \[F^k - {1\over \sqtw} k' (\vp \chi^k)
+ {i \over \sqtw} ( \psibar^m \sbar_m \chi^k) \]
\ddd \left. + Wk' \[ \fourth \ub - \fourth \tr (\lb \lb) + \third \Mb \ell
+{i \over 2} (\psibar^m \sbar_m \vp) \] \right\} + \hc
\label{spcf} \\
\ove \Lcal_{\stxt{VY}}^U &=& \sumC {b_\ap' \over 4}
\left\{
- {1 \over 2 \ua} (\La \La)_\ap
+ {\ua \over 4} k'' \vps
+ {\ua \over 2} G_{k;\ell} \ckl \right. \ddd
+ {1\over \sqtw} k' (\La_\ap \vp)
- \[ \Mbar + ( \psibar_m \sbar^{mn} \psibar_n ) \]
u_\ap \ln \( e^{-K/2} u_\ap / \mu^3 \)
\ddd
- k' u_\ap \[ \fourth \ub - \fourth \tr (\lb \lb) + \third \Mb \ell
+ {i \over 2} (\psibar^m \sbar_m \vp) \]
\ddd
- G_k \[\ua F^k + {i \over \sqtw} \ua ( \psibar^m \sbar_m \chi^k)
- (\La_\ap \chi^k) \]
\ddd
\left.
+ \ln \( e^{1-{K \over 2}} u_\ap / \mu^3 \)
\[F_{U_\ap} + {i \over \sqtw} ( \psibar^m \sbar_m \La_\ap) \]
\right\}
+ \hc
\label{vyuc} \\
\ove \Lcal_{\stxt{VY}}^\Pi &+& \ove \Lcal_{\stxt{thr}}^{\stxt{NP}} = \ddd
\sumC {1 \over 4}
\left\{
\ha \[ F_{\Ua} + {i \over \sqtw} ( \psibar^m \sbar_m \La_\ap)
- \ua \( \Mbar + ( \psibar_m \sbar^{mn} \psibar_n ) \) \]
\right. \ddd
+ \ha_k \[ \ua \Fk + {i \over \sqtw} \ua ( \psibar^m \sbar_m \chi^k)
- (\La_\ap \chi^k) \]
\ddd \left. - \half \ha_{k;\ell} \ua \ckl
\right\} + \hc
\label{vyth} \\
\ove \Lcal_{\stxt{thr}}^{\stxt{P}}
&=& \sumNC \fourth
\left\{
\hha \[\lla \( \Mbar + ( \psibar_m \sbar^{mn} \psibar_n ) \)
-\half f_\ap^{mn} f_{\ap mn} \right. \right.
\ddd
- {i \over 4} \e^{mnpq} f_{\ap mn} f_{\ap pq}
-2i (\la \s^m \D_m \lb)_\ap + \Dbf_\ap^2 - \Mb (\la \la)_\ap
\ddd
+ \( (\psi_m \s^{pq} \s^m \lb_\ap) + (\psib_m \sbar^{pq} \sbar^m \la_\ap) \)
\ddd
\qquad \times
\( i f_{\ap pq} + \half (\psi_p \s_q \lb_\ap)
+ \half (\psib_p \sbar_q \la_\ap) \)
\ddd \left.
- {i \over 2} \e^{mnpq} (\psib_m \sbar_n \la_\ap)
\( (\psi_p \s_q \lb_\ap) + (\psib_p \sbar_q \la_\ap) \) \]
\ddd
- \hha_k \[ \lla \(\Fk + {i \over \sqtw} ( \psibar^m \sbar_m \chi^k) \)
-i \sqtw (\chi^k \la_\ap) \Dbf_\ap
\right. \ddd \left.
+ \sqtw (\chi^k \s^{mn} \la^\ap) \hf_{\ap mn} \]
\left.
- \half \hha_{k;\ell} \ua \ckl
\right\} + \hc
\label{thrp} \\
\ove \Lcal_{\stxt{GS}}^X &=& {\dx \over 4} a_{X m} \left\{
2 B^m  + \tr(\lb \sbar^m \la)
+ 2i \[ (\vp \s^{nm} \psi_n) + (\psib_n \sbar^{nm} \vpb) \] \right.
\ddd
\left.
- 2\ell (\psib^n \sbar^m \psib_n) + i\ell \e^{mnpq} (\psib_n \sbar_p \psi_q)
\right\} + {\dx \over 4} \left\{ i \[ (\vp \la_X) - (\vpb \lb_X) \]
\right. \ddd \left.
+ \ell \[ (\psib^m \sbar_m \la_X) + (\lb_X \sbar^m \psi_m) \]
+ 2 \ell \Dbf_X \right\}
\label{gsvx}
\eeqa
Here we have evaluated $\Lcal_{\stxt{GS}}^X$ in
Wess-Zumino (WZ) gauge (see Eq.~\myref{wzd}).


\mysection{Expansion of $\Lcal_{\stxt{kin}}$}
\label{kine}
The kinetic Lagrangian $\Lcal_{\stxt{kin}}$,
defined in \myref{rar1}, is obtained
from the component expansion of $\hr_{\stxt{kin}}$,
defined in \myref{cac}.  With some effort the
following expressions are obtained:
\beqa
\hr_{\stxt{kin}} \| &=& -\eig f'' \vpbs
+ \eig f' \( u - (\la^\ap \la_\ap) + {4 \over 3} M \ell \)
\ddd
+ \sixth (2-f) M,
\label{kin0}
\\
\Dad \hr_{\stxt{kin}} \| &=& -\eig f''' \vp_\alpha \vpbs
+ \eig f'' \vp_\alpha \( u - (\la^\ap \la_\ap) + {4 \over 3} M \ell \)
\ddd
+ {1 \over 4 \sqtw} f' \La_\alpha
- {i \over 4} f' \la_\alpha^\ap \Dbf_\ap
+ \fourth f' (\s^{nm} \la^\ap)_\alpha \hf_{\ap nm}
\ddd
- \fourth f'' (\s^m \vpb)_\alpha B_m
+ {i \over 4} f'' (\s^m \vpb)_\alpha \Delta_m \ell
+ \sixth f'' \ell (\s^m \vpb)_\alpha b_m
\ddd
+ \fourth f'' \la_\alpha^\ap (\lb_\ap \vpb)
+ X_{0\alpha} \| + Z_\alpha \|
\ddd
+ 2 (\s^{cb})_\alpha^{\spc \gamma} T_{cb\gamma} \| .
\label{kin1}
\eeqa
We have left abbreviated $T_{cb\gamma} \|$, which is
given above in \myref{tor1}, as well as:
\beqa
Z_\alpha \| &=&
- \eig k''' \vpa \vpbs
- {i \over 4} k'' (\s^m \vpb)_\alpha \Delta_m \ell
- {i \over 4}k' \la_\alpha^\ap \Dbf_\ap
\ddd
+ \eig \( k'' \vpa + i k' (\s^m \psib_m)_\alpha \)
\( u - (\la^\ap \la_\ap) + {4 \over 3} M \ell \)
\ddd
+ \sixth k'' \ell (\s^m \vpb)_\alpha b_m
- \fourth k'' (\s^m \vpb)_\alpha B_m
- \fourth k'' \la_\alpha^\ap (\lb_\ap \vpb)
\ddd
+ \fourth k' (\s^{nm} \la^\ap)_\alpha \hf_{\ap nm}
+ {1 \over 4 \sqtw} k' \La_\alpha + \sixth k' M \vpa
+ \sixth k' (\s^m \vpb)_\alpha b_m
\ddd
- \iotw k' (\s^m \D_m \vpb)_\alpha
+ {i \over 4} k' (\s^m \sbar^n \psi_m)_\alpha
\( i \Delta_n \ell + {2 \over 3} \ell b_n - B_n \)
\ddd
+ {i \over 4} k' (\s^m \lb^\ap)_\alpha (\psi_m \la_\ap) ,
\label{zaex} \\
X_{0\alpha} \| &=&
{1 \over \sqtw} \Fbk \Gkkb \chiak
- {i \over \sqtw} \Gkkb (\s^m \chib^\kbar)_\alpha \Delta_m \pk
+ i G_k \la_\alpha^\ap(T_\ap \phi)^k .
\label{xzex}
\eeqa
These quantities also appear in the final piece
that contributes to $\Lcal_{\stxt{kin}}$:
\beqa
\lefteqn{ (\Ds \hr_{\stxt{kin}} + \hc)\| } && \nnn
&=& -\fourth f'''' \vps \vpbs
+ {2 \over 3}(f'' + \ell f''')(\vp \s^m \vpb) b_m
\ddd
+ \fourth f''' \[ \vpbs \( \ubar - (\lb^\ap \lb_\ap)
+ {4 \over 3} \Mbar \ell \) + \hc \]
\ddd
- f'''(\vp \s^m \vpb) B_m
+ f''' (\vp \la^\ap)(\vpb \lb_\ap)
+ {5 \over 6} f'' \[ M \vps + \hc \]
\ddd
+ f'' \[ i (\D_m \vp \s^m \vpb)
- {i \over 2} (\la^\ap \s^m \vpb) (\lb_\ap \psib_m)
\right. \ddd
+ {i \over 2} (\psib_m \sbar^n \s^m \vpb)
\( i \Delta_n \ell + B_n - {2 \over 3} \ell b_n \)
\ddd \left.
- {i \over 4} (\psi_m \s^m \vpb)
\( \ubar - (\lb^\ap \lb_\ap) + {4 \over 3} \Mbar \ell \)
+ \hc \]
\ddd
+ {1 \over \sqtw} f'' [(\vp \La) + \hc]
- f'' \Delta^m \ell \Delta_m \ell
\ddd
+ f'' \( B^m - {2 \over 3} \ell b^m \)
\( B_m - {2 \over 3} \ell b_m \)
- \half f' (F_U + \Fbar_\Ub)
\ddd
- \half f'' (\la^\ap \la^{(b)}) (\lb_\ap \lb_{(b)})
+f '' \( B_m -{2 \over 3} \ell b_m \) \tr(\la \s^m \lb)
\ddd
- \fourth f'' \nUc \nUcb
\ddd
+ f'' \[-i (\vp \la^\ap) \Dbf_\ap
+ (\vp \s^{nm} \la^\ap) \hf_{\ap nm} + \hc \]
\ddd
+ 2 {\cal R} - {4 \over 3} b^m b_m - {8 \over 3} M \Mb
\ddd
+ \( { k' + \ell k'' \over 1 - \third \ell k' } \)
\[ \vpau X_{0\alpha} \| + \vpau Z_\alpha \|
+ 2 (\vp \s^{cb})_{\alpha}^{\spc \gamma} T_{cb \gamma} \| + \hc \]
\ddd
+ \[ \Dau X_{0\alpha} \| + \Dau Z_\alpha \| + \hc \]
+ \eig f' [ \Ds \tWW \| + \hc ] .
\label{kin2}
\eeqa
Here we have abbreviated $\Dau X_{0\alpha} \|$ and $\Dau Z_\alpha \|$.
The quantity $\Dau X_{0\alpha} \|$ is obtained from
(BGG-4.2.13) with the replacement
$K(\phi,\phib) \to G(\phi,\phib)$.  Thus,
\beqa
\Dau X_{0\alpha} \| &=&
G_{k \kbar} \[ {\over} 2 \D^m \phik \D_m \phibk
-2 \Fk \Fbk - (\psi^m \chik) (\psib_m \chibk)
\right. \ddd
+ i (\chik \s^m \D_m \chibk) - i (\D_m \chik \s^m \chibk)
\ddd
+ i 2 \sqtw (\chik \la^\ap) (\phib T_\ap)^\kbar
- i 2 \sqtw (\chibk \lb^\ap) (T_\ap \phi)^k
\ddd
+ {i \over \sqtw} \Fbk (\psib_m \sbar^m \chik)
+ {i \over \sqtw} \Fk (\psi_m \s^m \chibk)
\ddd
+ {1 \over \sqtw} \( (\psib_m \sbar^n \s^m \chibk)
- 2 (\psib^n \chibk) \) \Delta_n \phik
\ddd \left.
+ {1 \over \sqtw} \( (\psi_m \s^n \sbar^m \chik)
- 2 (\psi^n \chik) \) \Delta_n \phibk \]
\ddd
- \half R_{k \kbar \ell \lbar} (\chik \chi^\ell)
(\chibk \chib^\lbar)
+ 2 G_k \Dbf^\ap (T_\ap \phi)^k .
\eeqa
After no small effort we obtain:
\beqa
\lefteqn{ (\Dau Z_\alpha + \hc)\| } && \nnn
&=&
- \fourth k'''' \vps \vpbs
- \half k' \( F_U + \hc \)
+ k'' \Delta^m \ell \Delta_m \ell
\ddd
+ 2 k' \D^m \Delta_m \ell
+ {1 \over 2 \sqrt{2}} k'' \[ (\La \vp) + \hc \]
+ \half k'' \Dbf^\ap \[ i (\vp \la_\ap)  + \hc \]
\ddd
+ \half k'' \hf^\ap_{nm} \[ (\vp \s^{nm} \la_\ap)  + \hc \]
- \half (k'' + k''') (\vp \la^\ap) (\vpb \lb_\ap)
\ddd
- \half k'' (\la^\ap \la^{(b)}) (\lb_\ap \lb_{(b)})
- \[k'' b_m + \half (k'' + k''') B_m \] (\vp \s^m \vpb)
\ddd
+ \[ \third (k' + 2 k'' \ell) b_m - k'' B_m \] (\la^\ap \s^m \lb_\ap)
\ddd
+ \( {2 \over 3} \ell b^m - B^m \)
\[ {4 \over 3} k' b_m + k'' \( {2 \over 3} \ell b^m - B^m \) \]
\ddd
- \fourth k'' \nUc \nUcb
\ddd
+ \left\{ \nUcb \[ -\sixth k' M + \fourth k''' \vpbs
\right. \right. \ddd \qquad \left. \left.
- {i \over 8} k'' (\psi_m \s^m \vpb) + \fourth k' (\psi^m \psi_m) \]
+ \hc \right\}
\ddd
+ \third k'' \[ M \vps + \hc \]
+ \half k'' \[ i (\D_m \vp \s^m \vpb) + \hc \]
\ddd
+ \fourth k'' \[ i (\psib_m \sbar^n \s^m \vpb)
\( i \Delta_n \ell + B_n - {2 \over 3} \ell b_n \) + \hc \]
\ddd
- \fourth k'' \[ (\la^\ap \s^m \vpb) (\lb_\ap \psib_m) + \hc \]
\ddd
+ k' (\psi^m \la^\ap)(\psib_m \lb_\ap)
- k' \[ (\psi^m \D_m \vp) + \hc \]
\ddd
- {1 \over 6} k' b_n \[i (\psi_m \s^n \sbar^m \vp) + \hc \]
+ {1 \over 6} k' \[ i \Mbar (\psi_m \s^m \vpb) + \hc \]
\ddd
+ k' (\psi^m \s^n \psib_m) \( B_n - {2 \over 3} \ell b_n \)
- \third k' b_m (\lb^\ap \sbar^m \la_\ap)
\ddd
+ \( { k' + 3 k'' \ell \over k' \ell -3 } \)
\[ \vpau X_{0\alpha} \| + \vpau Z_\alpha \| + \hc \]
\ddd
+ \( {k' (4 - k' \ell + k'' \ell^2) \over k' \ell - 3} \)
\[ (\vp \s^{cb})^\alpha T_{cb \alpha} \| + \hc \]
\ddd
+ \eig k' \[ \Ds \tWW \| + \hc \] .
\label{rar4}
\eeqa
Together with the notations defined in Appendix \ref{cna},
Eqs.~\myref{kin0}-\myref{rar4} provide the full component
expansion of $\Lcal_{\stxt{kin}}$.


\mysection{Expansion of $\Lcal_{\stxt{GS}}^0$}
\label{mgse}
A tedious calculation, using the methods described in
Appendix \ref{geid}, yields:
\beqa
\hr_{\stxt{GS}}^0\| &=&
\eig S \[ u - \tll \]
+ \half S_\kbar \[ \ell \Fbk
- {1 \over \sqtw} (\vpb \chibk) \]
\ddd
- {1 \over 4} \ell S_{\kbar;\lbar} \cbkl ,
\label{daz1} \\
\Dad \hr_{\stxt{GS}}^0\| &=&
\fourth S \[ {1 \over \sqtw} \La_\alpha
- i \la_\alpha^\ap \Dbf_\ap
+ (\s^{nm} \la^\ap)_\alpha \hf_{\ap nm} \]
\ddd
+ S_k \[ {1 \over 4 \sqtw} \chiak \( u - \tll \)
+ i \ell \la_\alpha^{\ap} (T_\ap \phi)^k \]
\ddd
+ \half S_\kbar \[ \Fbk \( \vpa - i \ell (\s^m \psib_m)_\alpha \)
- {1 \over \sqtw} \la_\alpha^\ap (\lb_\ap \chibk)
\right. \ddd
+ \Delta_m \pbk \( i (\s^m \vpb)_\alpha
+ \ell (\s^n \sbar^m \psi_n)_\alpha \)
\ddd \left.
+ \sqtw i \ell (\s^m \D_m \chib^\lbar)_\alpha
+ {1 \over \sqtw} (\s^m \chibk)_\alpha \( i \Delta_m \ell - B^m \) \]
\ddd
+ {1 \over \sqtw} S_{k \kbar} \chi_\alpha^k \[ \ell \Fbk
- {1 \over \sqtw} (\vpb \chibk) \]
-{1 \over 2 \sqtw} \ell S_{k \kbar;\lbar} \chiak \cbkl
\ddd
+ S_{\kbar;\lbar} \[
{i \over \sqtw} \ell \Delta_m \pbk (\s^m \chib^\lbar)_\alpha
- \fourth \vpa \cbkl \] ,
\label{daa}
\eeqa
\beqa
\Ds \hr_{\stxt{GS}}^0 \| &=&
- \half S F_U
+ S_k \[ -\half F^k \(u - (\la^\ap \la_\ap) \) + \half (\chik \La)
\right. \ddd
+ i (\vp \la^{\ap}) (T_\ap \phi)^k
+ \Dbf^\ap \( 2 \ell (T_\ap \phi)^k - {i \over \sqtw}(\chik \la^\ap) \)
\ddd \left.
+ \sqtw i \ell (\la^\ap T_\ap \chi)^k
+ {1 \over \sqtw} (\chik \s^{nm} \la^\ap) \hf_{\ap nm} \]
\ddd
+ S_\kbar \left\{
- \half \( \Fbk + {i \over \sqtw} (\psi_m \s^m \chibk) \)
\( \ub - (\lb^\ap \lb_\ap) + {4 \over 3} \Mb \ell \)
\right. \ddd
+ 2i \Delta_m \pbk \( i \Delta_m \ell - B_m \)
+ \sqtw i \D_m (\vp \s^m \chibk)
\ddd
- i \Fbk (\vp \s^m \psib_m)
+ \Delta_n \pbk (\vp \s^m \sbar^n \psi_m)
+ \sqtw \Mbar (\vpb \chibk)
\ddd
+ \half (\Lb \chibk)
- {i \over \sqtw} (\la^\ap \s^m \chibk) (\lb_\ap \psib_m)
+ {i \over \sqtw} \Dbf^\ap (\lb_\ap \chibk)
\ddd
+ {i \over \sqtw} (\psib_m \sbar^n \s^m \chibk)
\( i \Delta_n \ell + B_n - \ell b_n \)
\ddd
- {1 \over \sqtw} (\lb^\ap \sbar^{nm} \chibk) \hf_{\ap nm}
+ i \Delta_m \pbk (\la^\ap \s^m \lb_\ap)
\ddd
+ i (\phib T_\ap )^\kbar \[ (\vp \la^\ap) - 2 (\lb^\ap \vpb) \]
- 2 \ell \D^m \D_m \phib^k
\ddd
- {i \over 3 \sqtw} \ell \Mbar (\psi_m \s^m \chib^\kbar)
+ \sqtw \[ \D^m (\psib_m \chib^\kbar)
+ (\psib^m \D_m \chib^\kbar) \]
\ddd
- \ell (\phib T_\ap)^\kbar \[ (\psi_m \s^m \lb^\ap)
- (\psib_m \sbar^m \la^\ap) \]
\ddd
+ i \ell \Delta_n \phib^\kbar (\psi_m \s^n \psib^m)
- \ell \Fbar^\kbar (\psib^m \psib_m)
- {4 \over 3} \ell \Mbar \Fbk
\ddd
+ 2 \sqtw i \ell (\chib T_\ap \lb^\ap)^\kbar
+ 2 \sqtw i \ell \Gamma_{\lbar \mbar}^\kbar (\phib T_\ap)^\mbar
(\lb^\ap \chib^\lbar)
\ddd \left.
- R^\kbar_{\spc \mbar j \lbar} \[
{1 \over 2 \sqtw} (\vp \chi^j) (\chib^\mbar \chib^\lbar)
+ i \ell \Delta_n \phib^\mbar (\chi^j \s^n \chib^\lbar) \] \right\}
\ddd
+ S_{k \kbar} \left[ {\over}
\sqtw \Fbk (\vp \chik) + \sqtw F^k (\vpb \chibk) - 2 \ell \Fk \Fbk
\right. \ddd
- \sqtw i \ell (\chik \la^\ap) (\phib T_\ap)^\kbar
+ \half \ell R^\kbar_{\spc \mbar p \lbar} (\chik \chi^p)
(\chib^\mbar \chib^\lbar)
\ddd
- \sqtw \ell \Delta_n \phi^k (\psib_m \sbar^n \s^m \chib^\kbar )
+ (\chik \s^m \chibk)
\(i \Delta_m \ell + {2 \over 3} \ell b_m - B_m \)
\ddd \left.
- (\chik \la^\ap) (\chibk \lb_\ap)
+ \sqtw i \Delta_m \pbk (\chi^k \s^m \vpb) {\over} \right]
\ddd
+ S_{k;\ell} \[ \fourth \ckl \( u - (\la^\ap \la_\ap) \)
+ \sqtw i \ell (\chi^\ell \la^\ap) (T_\ap \phi)^k \]
\ddd
+ S_{\kbar; \lbar} \[
\fourth (\chib^\kbar \chib^\lbar)
\( \ubar - (\lb^\ap \lb_\ap) - {4 \over 3} \Mbar \ell \)
\right. \ddd
+ \sqtw i \Delta_m \phibk (\vp \s^m \chib^\lbar)
- 2 \ell \Delta_m \phib^\kbar \Delta^m \phib^\lbar
\ddd \left.
+ 2 \sqtw i \ell (\lb^\ap \chib^\kbar) (\phib T_\ap)^\lbar
{\over} \]
+ S_{k \kbar; \lbar} \[ -{1 \over \sqtw} \cbkl (\vp \chik)
\right. \ddd \left.
+ \ell \Fk (\chibk \chib^\lbar)
+ 2i \ell \Delta_m \phib^\lbar (\chi^k \s^m \chib^\kbar)
{\over} \]
\ddd
+ S_{k \lbar; \ell} \[ -{1 \over \sqtw} \ckl (\vpb \chib^\lbar)
+ \ell \Fbar^\lbar (\chik \chi^\ell) \]
\ddd
- \half \ell S_{k \kbar; \ell ; \lbar} (\chik \chi^\ell) (\chibk \chib^\lbar)
+ \eig S \Ds \tWW\|
\ddd
+ 2 \sqtw \ell S_\kbar (\chibk \sbar^{cb})_\adot T_{cb}^{\bspc \adot} \| .
\label{dab}
\eeqa


\end{document}